\documentclass{vgtc}                          

\ifpdf
  \pdfoutput=1\relax                   
  \usepackage{graphicx}                
  \DeclareGraphicsExtensions{.pdf,.png,.jpg,.jpeg} 
\else
  \usepackage{graphicx}                
  \DeclareGraphicsExtensions{.eps}     
\fi%

\graphicspath{{figures/}{pictures/}{images/}{./}} 

\usepackage{microtype}                 
\PassOptionsToPackage{warn}{textcomp}  
\usepackage{textcomp}                  
\usepackage{mathptmx}                  
\usepackage{times}                     
\usepackage{cite} 
\usepackage{tabu}                      
\usepackage{booktabs}                  
\usepackage{float}
\usepackage{wrapfig}

\usepackage[usenames,dvipsnames,table]{xcolor}
\usepackage{color}
\usepackage{graphicx}
\usepackage{capt-of}

\usepackage{wasysym}

\usepackage{amssymb}

\usepackage{xspace}

\usepackage{enumitem}

\usepackage{hyperref}

\usepackage{tabu}

\usepackage{picinpar}

\usepackage[utf8]{inputenc}

\usepackage{microtype}
\usepackage{wrapfig}

\usepackage{subcaption}

\usepackage{hyphenat}
\usepackage[english]{babel}
\addto\extrasenglish{%
}

\usepackage{xspace}
\newcommand{\SIM}{\texttt{SIM}\xspace} 
\newcommand{\DIS}{\texttt{DIS}\xspace} 

\newcommand{\noClutter}{\texttt{0C}\xspace} 
\newcommand{\oneClutter}{\texttt{1C}\xspace} 
\newcommand{\twoClutter}{\texttt{2C}\xspace} 

\usepackage[normalem]{ulem}

\catcode`\@ = 11
\newdimen\@InsertBoxMargin
\newcount\@numlines    
\newcount\@linesleft   
\def\ParShape{%
    \@numlines = 0
    \def\@parshapedata{ }
    \afterassignment\@beginParShape
    \@linesleft
}%
\def\@beginParShape{%
    \ifnum \@linesleft = 0
      \let\@whatnext = \@endParShape
    \else
      \let\@whatnext = \@readnextline
    \fi
    \@whatnext
}%
\def\@endParShape{%
    \global\parshape = \@numlines \@parshapedata
}%
\def\@readnextline#1 #2 #3 {
    \ifnum #1 > 0
      \bgroup  
        \dimen0 = \hsize
        \advance \dimen0 by -#2  
        \advance \dimen0 by -#3  
        \count0 = 0
        \loop
          \global\edef\@parshapedata{%
            \@parshapedata    
            #2                
            \space            
            \the\dimen0       
            \space            
          }%
          \advance \count0 by 1
          \ifnum \count0 < #1
        \repeat
      \egroup
      \advance \@numlines by #1
    \fi
    \advance \@linesleft by -1
    \@beginParShape
}%
\newbox\@boxcontent     
\newcount\@numnormal    
\newdimen\@framewidth   
\newdimen\@wherebottom  
\newif\if@byframe       
\@byframefalse
\def\InsertBoxC#1{%
  \leavevmode
  \vadjust{
    \vskip \@InsertBoxMargin
    \hbox to \hsize{\hss#1\hss}
    \vskip \@InsertBoxMargin
  }%
}%
\def\InsertBoxL#1#2{%
  \@numnormal = #1
  \setbox\@boxcontent = \hbox{#2}%
  \let\@side = 0
  \futurelet \@optionalparameter \@InsertBox
}
\def\InsertBoxR#1#2{%
  \@numnormal = #1
  \setbox\@boxcontent = \hbox{#2}%
  \let\@side = 1
  \futurelet \@optionalparameter \@InsertBox
}%
\def\@InsertBox{%
  \ifx \@optionalparameter [
    \let\@whatnext = \@@InsertBoxCorrection
  \else
    \let\@whatnext = \@@InsertBoxNoCorrection
  \fi
  \@whatnext
}%
\def\@@InsertBoxCorrection[#1]{%
  \ifx \@side 0
    \@@InsertBox{#1}{0}{{\the\@framewidth} 0cm}%
  \else
    \@@InsertBox{#1}{1}{0cm {\the\@framewidth}}%
  \fi
}%
\def\@@InsertBoxNoCorrection{%
  \@@InsertBoxCorrection[0]%
}%
\def\@@InsertBox#1#2#3{%
  \MoveBelowBox
  \@byframetrue
  \@wherebottom = \baselineskip
  \multiply \@wherebottom by \@numnormal
  \advance \@wherebottom by 2\@InsertBoxMargin
  \advance \@wherebottom by \ht\@boxcontent
  \advance \@wherebottom by \pagetotal
  \ifdim \pagetotal = 0cm
    \advance \@wherebottom by -\baselineskip  
  \fi
  \advance \@wherebottom by #1\baselineskip
  \@framewidth = \wd\@boxcontent
  \advance \@framewidth by \@InsertBoxMargin
  \bgroup  
    \ifdim \pagetotal = 0cm
      \dimen0 = \vsize
    \else
      \dimen0 = \pagegoal
    \fi
    \ifdim \@wherebottom > \dimen0
      \immediate\write16{+--------------------------------------------------------------+}%
      \immediate\write16{| The box will not fit in the page. Please, re-edit your text. |}%
      \immediate\write16{+--------------------------------------------------------------+}%
      \vrule width \overfullrule
    \fi
  \egroup
  \prevgraf = 0
  \vbox to 0cm{%
    \dimen0 = \baselineskip
    \multiply \dimen0 by \@numnormal
    \advance \dimen0 by -\baselineskip
    \setbox0 = \hbox{y}%
    \vskip \dp0
    \vskip \dimen0
    \vskip \@InsertBoxMargin
    \ifnum #2 = 1
      \vtop{\noindent \hbox to \hsize{\hss \box\@boxcontent}}%
    \else
      \vtop{\noindent \box\@boxcontent}%
    \fi
    \vss
  }%
  \vglue -\parskip
  \vskip -\baselineskip
  \everypar = {%
    \ifdim \pagetotal < \@wherebottom
      \bgroup  
        \dimen0 = \@wherebottom
        \advance \dimen0 by -\pagetotal
        \divide \dimen0 by \baselineskip
        \count1 = \dimen0
        \advance \count1 by 1
        \advance \count1 by -\@numnormal
        \ifnum #2 = 1
          \ParShape = 3
                      {\the\@numnormal}   0cm   0cm
                      {\the\count1}       0cm   {\the\@framewidth}
                      1                   0cm   0cm
        \else
          \ParShape = 3
                      {\the\@numnormal}   0cm                  0cm
                      {\the\count1}       {\the\@framewidth}   0cm
                      1                   0cm                  0cm
        \fi
      \egroup
    \else
      \@restore@    
    \fi
  }%
  \def\par{%
      \endgraf
      \global\advance \@numnormal by -\prevgraf
      \ifnum \@numnormal < 0
        \global\@numnormal = 0
      \fi
      \prevgraf = 0
  }%
}%
\def\MoveBelowBox{%
  \par
  \if@byframe
    \global\advance \@wherebottom by -\pagetotal
    \ifdim \@wherebottom > 0cm
      \vskip \@wherebottom
    \fi
    \@restore@
  \fi
}%
\def\@restore@{%
    \global\@wherebottom = 0cm
    \global\@byframefalse
    \global\everypar = {}%
    \global\let \par = \endgraf
    \global\parshape = 1 0cm \hsize
}%
\ifx \documentclass \@Dont@Know@What@It@Is@
\else
  \let \pageno = \c@page
\fi

\catcode`\@ = 12

\newcommand{\subhead}[1]{\vspace{2pt} \noindent \textbf{#1}}

\onlineid{0}

\vgtccategory{Research}

\vgtcinsertpkg

\title{Evaluating Ordering Strategies of Star Glyph Axes}

\newcommand\acc[1]{\hspace{0.02cm}\hspace{-0.04cm}#1}

\author{Matthias Miller\thanks{e-mail: firstname.lastname@uni-konstanz.de}\\ %
       \parbox{1.4in}{\scriptsize \centering University of Konstanz }
\and Xuan Zhang\thanks{e-mail: xuan.zhang2@rwth-aachen.de}\\ 
     \parbox{1.4in}{\scriptsize \centering RWTH Aachen University }
\and Johannes Fuchs\acc{$^*$}\\ %
     \parbox{1.4in}{\scriptsize \centering University of Konstanz }
\and Michael Blumenschein\acc{$^*$}\\ %
     \parbox{1.4in}{\scriptsize \centering University of Konstanz }
}


\abstract{
Star glyphs are a well-researched visualization technique to represent multi-dimensional data.
They are often used in small multiple settings for a visual comparison of many data points.
However, their overall visual appearance is strongly influenced by the ordering of dimensions.
To this end, two orthogonal categories of layout strategies are proposed in the literature: order dimensions by \emph{similarity} to get homogeneously shaped glyphs vs. order by \emph{dissimilarity} to emphasize spikes and salient shapes.
While there is evidence that salient shapes support clustering tasks, evaluation, and direct comparison of data-driven ordering strategies has not received much research attention. 
We contribute an empirical user study to evaluate the efficiency, effectiveness, and user confidence in visual clustering tasks using star glyphs.
In comparison to similarity-based ordering, our results indicate that dissimilarity-based star glyph layouts support users better in clustering tasks, especially when clutter is present.

} 

\begin{document}



\firstsection{Introduction}
\maketitle

Data glyphs are compact visual representations of multi-dimensional data points.
Due to their small graphical appearance, they can be used in various settings like within node-link diagrams~\cite{DBLP:conf/vda/Erbacher02}, treemaps~\cite{DBLP:conf/vissym/00010M12}, tables~\cite{DBLP:conf/vizsec/KintzelFM11}, or geographic maps~\cite{DBLP:journals/tvcg/FuchsIBFB14}. 
For instance, star glyphs are employed in the medical domain~\cite{DBLP:conf/simvis/RopinskiP08} and can be used to show the spatial distribution of food production~\cite{opach_star_2018}.

Due to their use of visual variables, star glyphs~\cite{Siegel:1972} are an adequate choice to encode single data points comprising numerical data. 
The glyph's axes represent the data dimensions, and their lengths encode numeric values. 
Since glyphs are versatile, different design variations of star glyphs emerged in literature.
Many have already been extensively analyzed by the community (e.g., \cite{DBLP:journals/tvcg/FuchsIBFB14}, see \cite{DBLP:journals/tvcg/FuchsIBK17} for a full enumeration).
However, there is not much empirical research about the effect of axes ordering strategy on visual comparison tasks.

The ordering influences the shape of a star glyph and affects its readability and similarity judgment~\cite{klippel_star_2009, DBLP:journals/cartographica/KlippelHLW09}. 
Therefore, we need (task-based) guidelines to arrange the dimensions in star glyphs~\cite{ward_multivariate_2008}. 

Numerous ordering strategies for star glyphs have been proposed 
\cite{DBLP:conf/infovis/AnkerstBK98,DBLP:conf/iv/ArteroOL06,DBLP:conf/infovis/PengWR04,ward_multivariate_2008,DBLP:journals/csda/FriendlyK03,DBLP:conf/infovis/YangPWR03,klippel_star_2009, DBLP:journals/cartographica/KlippelHLW09,KAYAERT2009708}
which can be grouped into \emph{similarity-based} (short: \SIM), favoring homogeneous shapes, and \emph{dissimilarity-based orderings} (short: \DIS), emphasizing spikes and salient shapes. 
Some approaches also discuss symmetry, monotonicity, convexity and concavity, feature saliency, and user-driven relationships among neighboring dimensions. 
The ordering strategies typically analyze the relationship among all pair-wise dimensions and then adjust the axes of every star glyph simultaneously according to a metric (e.g., \SIM or \DIS). 
However, this also means that not all glyphs will result in the desired shape. 
In particular, outliers may be encoded by shapes which the reordering algorithm is trying to avoid.

We address the research question: ``Which ordering strategy is most useful for similarity search and data grouping tasks (clustering) using star glyphs?''. 
According to the task taxonomy by Andrienko and Andrienko~\cite{DBLP:books/daglib/0015278}, similarity search, and grouping are among the most common analysis tasks for glyphs~\cite{DBLP:journals/tvcg/FuchsIBFB14}. 
While different strategies have been proposed, they are not yet evaluated by empirical studies. 
Klippel et al.~\cite{klippel_star_2009, DBLP:journals/cartographica/KlippelHLW09} evaluated the influence of a star glyph's shape in grouping tasks. 
Although they found out that salient shapes, e.g., having spikes, can support grouping tasks, they did not apply a dimension ordering strategy that considers these salient properties.

Sorting the dimensions by dissimilarity favors the spikey-design, which Klippel et al. states to be promising for grouping. 
We compare this ordering strategy with the similarity-based design which is often proposed in the literature \cite{ward_multivariate_2008,DBLP:journals/csda/FriendlyK03,borg_staufenbiel_snowflakes_1992, DBLP:conf/infovis/AnkerstBK98}. 
We conducted an empirical user study with 15 participants to evaluate the efficiency, data clustering quality, noise identification quality, and user confidence between the two different strategies (first independent variable). 
Our results show that star glyphs, ordered by a dissimilarity-based layout, support users better in a clustering task. 

Real-world data often contains non-relevant dimensions with clutter and noise that may distort interesting patterns~\cite{DBLP:series/isrl/2015-72}.
Additionally, clusters do often not span across all dimensions but exist only in subspaces~\cite{DBLP:journals/tkdd/KriegelKZ09}. 
Therefore, we investigate  impact of clutter on cluster identification and reordering strategies as a second independent variable. 
We use the term \textit{clutter dimensions} to describe attributes that do not discriminate clusters but hinder the comprehension of feature relationships in the data~\cite{DBLP:conf/infovis/PengWR04}.
Therefore, we also investigate the influence of clutter separately, and in combination with the ordering strategies. 
For replicability and reproducibility, the material of the study (benchmark data, study results, analysis scripts, and code) is publicly available at \href{https://osf.io/bje89}{https://osf.io/bje89}. 
\section{Related Work}
Finding an optimal star glyph axes ordering has proven to be NP-complete~\cite{DBLP:conf/infovis/AnkerstBK98} and more research is required~\cite{ward_multivariate_2008,rzezniczak2013evaluation}. 
It is related to the ordering of axes in parallel coordinates~\cite{DBLP:conf/visualization/InselbergD90,DBLP:conf/apvis/ZhangMM12,DBLP:conf/infovis/AnkerstBK98,DBLP:journals/tvcg/DasguptaK10,DBLP:journals/tvcg/TatuAEBTMK11}, RadViz~\cite{DBLP:conf/visualization/HoffmanGMGS97,cheng_radviz_2017,DBLP:conf/ieeevast/AlbuquerqueELTM10,DBLP:conf/pakdd/CaroFF10}, ArcViz~\cite{DBLP:conf/kse/Long18}, and other axes-based radial visualizations as summarized by Behrisch et al.~\cite{DBLP:journals/cgf/BehrischBKSEFSD18}. 
Ordering algorithms typically define an objective function, modeling a good dimension order (according to their interpretation) and apply a heuristic to find an axes order which maximizes the objective function~\cite{ward_multivariate_2008}. 

\subsection{Dimension Ordering Strategies}
Different visual characteristics can be subject to shape optimization when applying specific ordering strategies of the star glyph axes.
Ward~\cite{ward_multivariate_2008} summarizes four major strategies which have been extended by others: \emph{user- and data-driven}, \emph{correlation- and similarity-driven}, \emph{spikes and salient shapes}, and \emph{symmetry-driven}.  

\subhead{User-driven} dimension orderings enable experts to adjust the shape of a star glyph based on their domain knowledge~\cite{DBLP:journals/tvcg/SachaSSKEK14}. 
Users can select a data point to sort the data dimensions with ascending or descending order (\emph{data-driven}) to reveal patterns between records~\cite{ward_multivariate_2008}.

\InsertBoxR{0}{\parbox{0.9cm}{\includegraphics[width=\linewidth]{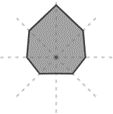}}}[0]\subhead{Correlation- and similarity-driven} strategies improve star glyphs by adjacent placement of similar axes to support understanding of clusters, outliers, and  relationships~\cite{DBLP:conf/eurographics/BorgoKCMLHWC13}.

\noindent Ankerst et al. propose heuristic algorithms based on similarity  for star glyphs to improve the overall perception~\cite{DBLP:conf/infovis/AnkerstBK98}. Similarly, Artero et al. use similarity heuristics of attributes to apply dimension-ordering and take perceptional aspects as Gestalt Laws into account by applying dimension reduction~\cite{DBLP:conf/iv/ArteroOL06}. Yang et al. combine similarity-based ordering with a hierarchical structure of the dimension to enable interactive exploration of high-dimensional subspaces~\cite{DBLP:conf/infovis/YangPWR03}. Friendly and Kwan argue that using correlation-based ordering in star glyphs supports the identification of shape irregularities~\cite{DBLP:journals/csda/FriendlyK03}. The authors did not conduct a survey to underpin their statement.

\InsertBoxR{0}{\parbox{0.8cm}{\includegraphics[width=\linewidth]{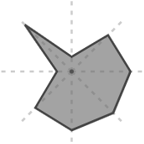}}}
\subhead{Spikes and salient shapes} such as ``\textit{has-one-spike}'' are helpful in visual grouping tasks of data points according to Klippel et al~\cite{DBLP:journals/cartographica/KlippelHLW09}. They argue, that concavity is more suitable for comparability than convexity, which is especially true for the ``star"

\InsertBoxR{0}{\parbox{0.8cm}{\includegraphics[width=\linewidth]{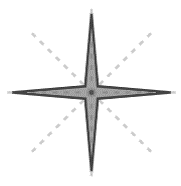}}}[0] \noindent
glyph, due to the large variations between adjacent dimensions. The salience of dissimilar neighboring axes shall enhance the comparison speed and help to detect changes.

\noindent Klippel et al. showed that the star glyph shape with eight dimensions influences classification tasks~\cite{klippel_star_2009}. Especially, in contrast to earlier work that state that similarity-driven orderings improve high-dimensional visualizations, dissimilarity between neighboring axes contribute \textit{salient} properties that are perceptually more noticeable. 

\InsertBoxR{0}{\parbox{1.48cm}{\includegraphics[width=\linewidth]{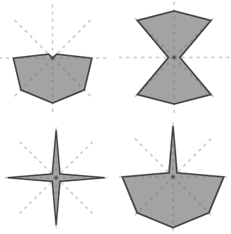}}}[1]
\subhead{Symmetry-driven} reordering methods help to reduce the visual complexity of star glyphs and, therefore, support  comparison tasks by improving memorability~\cite{KAYAERT2009708}. By providing some examples, Peng et al. argue that orderings with simple and symmetric as well as monotonic shapes of the star glyphs facilitate the identification of value differences between multiple dimensions~\cite{DBLP:conf/infovis/PengWR04}. 
They emphasize that symmetry and similarity are primary factors to identify patterns. 
For this, Gestalt Laws are a solid foundation for perception design~\cite{colin2004information}. Peng et al. state that star glyphs can be optimized by aligning the symmetry on the vertical or horizontal axis~\cite{DBLP:conf/infovis/PengWR04}. An additional rotation optimization step can be included in the pipeline to find the best global rotation for all star glyphs of a dataset. 
Rotation can be applied, e.g., on top of similarity or spike-based ordering.

\subsection{Empirical Studies and Research Gap} 
While many ordering strategies, algorithms, and heuristics have been proposed star glyph dimension ordering, empirical evaluation is missing. 
Previous approaches mainly argue by showing examples or providing arguments w.r.t. to e.g., Gestalt laws. 
While this is useful to find differences between strategies, we also need empirical evidence to directly compare strategies respecting scalability, performance, analysis tasks, data characteristics, and user perception~\cite{ward_multivariate_2008}.

We are only aware of two studies conducted by Klippel et al.~\cite{klippel_star_2009,DBLP:journals/cartographica/KlippelHLW09}. 
Their results indicate that spikes and salient shapes have a positive effect on visual grouping tasks and colored axes positively affect the processing speed and reduce the negative influence of shape saliency on rotated data glyphs. 
However, Klippel et al. did not directly compare different ordering strategies or evaluated them against a benchmark. 
Instead, they designed different star glyph shapes and analyzed how participants grouped them by their understanding of similarity during an exploratory analysis task. 
In our study, we aim to close this research gap by comparing two proposed reordering strategies using a controlled, empirical user experiment.
\begin{figure}[h]
    \centering
    \includegraphics[width=\linewidth]{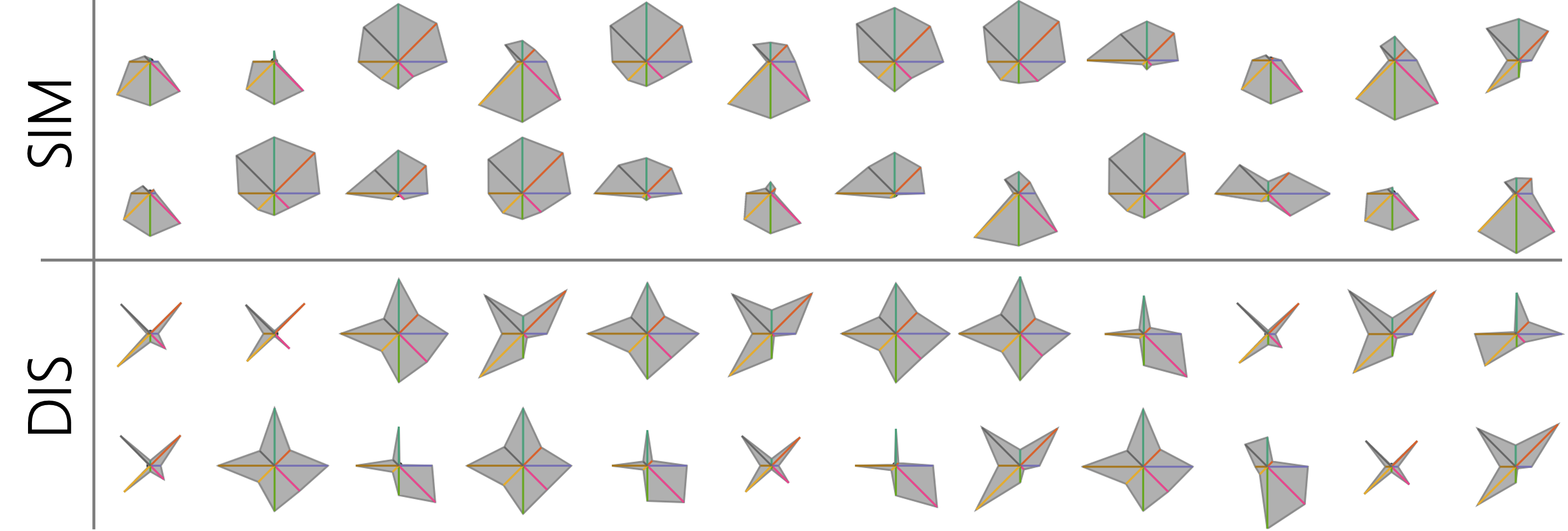}
    \caption{\textbf{Comparison of similarity (\SIM) and dissimilarity (\DIS) based ordering} using the same data records.}
    \label{fig:sg-example-comparison}
\end{figure}

\section{Empirical User Study}
We evaluate whether a similarity- (homogeneous shape, short: \SIM) or dissimilarity-based layout (spike and salient shape, short: \DIS) is more efficient and effective for a visual clustering task.  
We designed our study based on Klippel et al.'s work~\cite{klippel_star_2009,DBLP:journals/cartographica/KlippelHLW09}.
We adopted the task, user interface, glyph design (including colored axes), and datasets' dimensionality (eight dimensions). 
However, in contrast, we applied two different reordering algorithms (\SIM and \DIS) and different clutter levels as independent factors, and evaluate the results using a benchmark dataset.

\subsection{Experimental Design and Hypotheses}
The participants had to manually assign star glyphs into reasonable clusters and identify noise, i.e., items not belonging to any cluster. 
In the study, we used the term \emph{group} instead of cluster. 
To assess the performance, we use four dependent variables as quality measures: (i) \emph{task completion time}, (ii) \emph{quality of groups}, (iii) \emph{quality of identified noise}, and (iv) the \emph{confidence} of the participants.

\subhead{Participants.} 
We recruited 15 participants from the local student population (seven female, two bachelor, twelve master, one PhD student).
The age ranged from 20 -- 27 years with a median of 23. 
The participants had a different background in data analysis and visualization: ten had general knowledge in data analysis, four had data visualization experience, and one has used star glyphs before. 
All participants received a compensation of 10 EUR.

\InsertBoxR{0}{\parbox{0.12\linewidth}{\includegraphics[width=\linewidth]{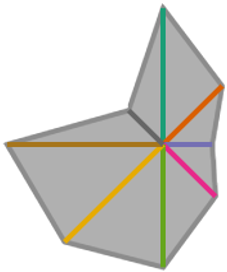}}}[1]
\subhead{Glyph Design and Implementation.}
The glyphs are designed analog to Klippel et al.'s work~\cite{klippel_star_2009,DBLP:journals/cartographica/KlippelHLW09} using a contour, gray background, and colored axes. 
We used ColorBrewer~\cite{harrower2003colorbrewer} to select diverging colors and applied the ordering algorithm by Ankerst et al.~\cite{DBLP:conf/infovis/AnkerstBK98}. 
The Euclidean distance is used to measure the (dis)similarity between dimensions. 
We ran an exhaustive search to find the permutation with the highest (\SIM) and lowest (\DIS) similarity. 
An example of star glyphs with the two orderings is depicted in Fig.~\ref{fig:sg-example-comparison}.
Orientation (rotation) of the star glyphs is not considered and chosen randomly.
All orderings are pre-computed not to influence the run time during the study. 
We provide the study and the ordering strategy implementation on our websites\footnote{\url{http://subspace.dbvis.de/sg-study} and \url{*/sg-ordering}.}.

\subhead{Hypotheses.}
We address the following two hypotheses: 

\noindent \textbf{H1.} 
Clutter negatively influences visual comparison. 
With increasing clutter, the performance of grouping tasks drops, independent of the axes ordering. 
In particular, we expect users to be \textbf{(a)}~slower, \textbf{(b)}~less accurate in grouping accuracy, \textbf{(c)}~less accurate in noise identification, and \textbf{(d)}~less confident of their grouping.

\noindent \textbf{H2.} 
Klippel et al.~\cite{klippel_star_2009,DBLP:journals/cartographica/KlippelHLW09} argue that spikes and salient shapes support users in similarity estimation and grouping tasks. 
Therefore, the performance of users should increase with a dissimilarity-based ordering. 
Furthermore, we hypothesize that the salient shapes should support users even more if the data contains clutter since sharp edges are more perceptually apparent. 
In particular, we expect users to be \textbf{(a)}~faster, \textbf{(b)}~more accurate in grouping accuracy, \textbf{(c)}~more accurate in noise identification, and \textbf{(d)}~more confident of their grouping when dimensions are ordered by dissimilarity.

\subsection{Benchmark Datasets}
We manually created 18 different datasets using the PCDC tool~\cite{DBLP:conf/vissym/BremmHLF12}.
Every dataset contains 50 records of which 2--7 data points are selected as \emph{noise} (randomly distributed across all dimensions).
The remaining data points are grouped into 2, 3, or 4 clusters with similar cluster sizes. 
Besides, we introduced \emph{clutter dimensions} which do not discriminate any cluster, since we uniformly distributed all data points across the clutter dimensions.
6 datasets contain no clutter (\noClutter), 6 one- (\oneClutter), and 6 two clutter dimensions (\twoClutter). For instance, in condition \twoClutter a dataset consists of six dimensions discriminating the cluster, while the remaining two introduce clutter. Thus, we generated the datasets to keep the number of dimensions consistent.
To verify the manually created clusters, we run a DBSCAN~\cite{DBLP:conf/kdd/EsterKSX96} (parameters: $minPts=3$, $\epsilon = 0.5$) on all datasets.

\begin{figure}[t]
    \centering
    \includegraphics[width=\linewidth]{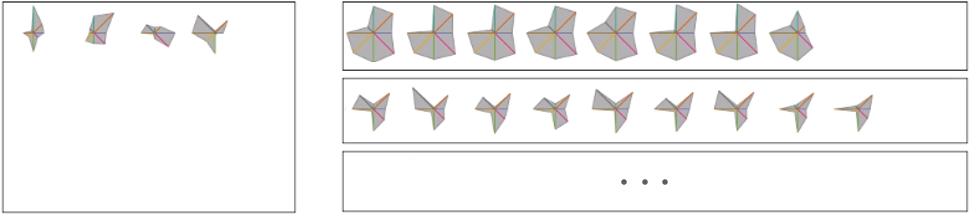}
    \caption{\textbf{Study prototype.} Users can group visually similar star glyphs using drag\&drop. Noise points remain in the left panel. 
    }
    \label{fig:sg-user-interface}
\end{figure}

\subsection{Tasks, Procedure, and Data Analysis}

\subhead{Tasks and Procedure.} 
Each study took an hour on average. 
Participants filled out a consent form, demographics, and report on previous knowledge in data analysis, information visualization, and star glyphs. 
Afterward, we described how to read the visual encoding of a star glyph using an artificial car dataset as an example. 
Specifically, we clarified that star glyphs with similar shapes on different axes are not similar (rotation invariance).
Finally, the participants performed three training trials before the study was recorded. 

To conduct the study, we used a 27-inch screen with 2560x1440 resolution and a mouse to execute given tasks. 
Every participant had to perform 18 trials, leading to $15\ \texttt{participants} \times 18\ \texttt{trials} = 270\ \texttt{trials}$ for the entire study.  
In between two trials, we showed a blank screen with the term `break' to motivate the participants to have regular breaks. 
Each trial consisted of manually grouping all 50 star glyphs of one dataset into distinct groups and noise. Then, the participants stated the confidence level about their selection on a 7-point Likert-scale. 
We did not provide the number of clusters per dataset and explicitly told the participants that there might be glyphs which do not belong to any group (noise). 

\autoref{fig:sg-user-interface} shows our interface. 
Participants were able to add new or delete groups. 
Glyphs can be interactively assigned to groups by drag\&drop. If a glyph was considered to be noise, then it remained in the left panel. Participants were able to undo or change a grouping also using drag\&drop. 
In the study, we did not constrain the task completion time. 
We ended the study with an interview about the participants' strategy and preferences regarding the \SIM and \DIS ordering by showing examples. 
Questions and answers were recorded.

\subhead{Randomization.}
Each participant performed 18 trials, i.e., the grouping task on all benchmark datasets was equally distributed between \SIM and \DIS.
We randomized the order of the trials as follows: 
First, we grouped the datasets into their level of difficulty based on the amount of clutter (\noClutter, \oneClutter, \twoClutter). 
Then, participants performed the trials with increasing difficulty, i.e., 6~\texttimes~\noClutter, then 6~\texttimes~\oneClutter, and finally 6~\texttimes~\twoClutter. 
For every clutter condition, we randomized the dataset order and randomly assigned 3~\texttimes~\SIM and 3~\texttimes~\DIS. 
We attached the randomization algorithm and our configuration in the supplementary material. 
A summary of our trials: 
\begin{center}
    \vspace*{-0.2cm}
	\begin{tabular}{rll}
	  	3 & levels of difficulty (clutter: \noClutter, \oneClutter, \twoClutter ) & \texttimes \\
	  	2 & ordering strategies  (\SIM, \DIS ) & \texttimes \\
	  	3 & trials  (2, 3, 4 clusters) & \texttimes \\
	  	15 & participants & = \\
		\hline
    	\textbf{270} & \textbf{trials in total} \\
	\end{tabular}
    \vspace*{-0.2cm}
\end{center}

\subhead{Data Collection, Post-Processing, and Analysis.}
In each trial, we recorded the grouping task completion time, the selected groups and noise, and the participants' confidence.
Some participants created groups with only one or two glyphs. 
Thus, in a post-processing step, we converted such small groups into noise to execute a more coherent analysis.
We measured the quality of the identified noise by computing the Jaccard index between noise and ground truth noise. 

\noindent The grouping quality is also based on the Jaccard index between the grouping and ground truth. 
However, since participants could have also selected too few or too many groups, we structured our quality computation as a two-step process: 
First, we computed the average Jaccard index of each group to its best match in the ground truth.
Second, we computed the average Jaccard index of every ground truth cluster to its best match in the selection. 
Using this method, we considered too few, too many groups, as well as too few and too many records per group.
The final clustering quality is the average score of both steps.

\subsection{Results and Statistical Analysis}
We executed a statistical analysis to summarize the study results. 
We report all statistically significant findings ($p < .05$) and some interesting trends visible in the data. 
We check for normal distribution using a \emph{one-sample Kolmogorov-Smirnov test}. 
For a better comparison, we always report the median ($\bar{x}$) and, additionally, the mean ($\mu$) for normally distributed samples. 
Analysis scripts and detailed results can be found in the supplementary material. 

\subhead{Statistical tests.}
\emph{Confidence} is measured as Likert-scale. 
Therefore, a \emph{Pearson's Chi-squared test} is used for the analysis.
Given the non-normal nature of the measures \emph{time}, \emph{cluster quality}, and \emph{noise identification quality} w.r.t. \noClutter, \oneClutter, and \twoClutter, we used a \emph{non-parametric Friedman's test} and a \emph{Wilcoxon signed rank test with Bonferroni correction} (Post-hoc). 
The same measures do also not follow a normal distribution w.r.t. the strategies \SIM and \DIS. 
Hence we used a \emph{Wilcoxon signed rank test with continuity correction}. 
Considering \SIM and \DIS within \oneClutter and \twoClutter reveal normal distributed samples for the measures \emph{time}, \emph{quality of clustering}, and \emph{quality of noise identification}. 
Hence, we use a \emph{paired t-test} for the statistical analysis.

\subhead{Task Completion Time.}

\noindent \textbf{H1a.} Task completion time increased with \emph{clutter levels}, but not significantly: 
\noClutter ($\bar{x} = 162.5s$), \oneClutter ($179.5s$), and \twoClutter ($184.0s$). 

\noindent \textbf{H2a.} Using the \emph{ordering} \DIS ($176.0s$) users completed the grouping task slightly faster than \SIM ($178.0s$), but only for datasets with clutter dimensions. 
\noClutter: \DIS ($168.0s$) vs. \SIM ($158.0s$),
\oneClutter: $180.0s$ / $179.0s$, and 
\twoClutter: $180.0s$ / $190.0s$. 
Differences are not significant.

\begin{figure}[t]
    \centering
    \includegraphics[width=\linewidth]{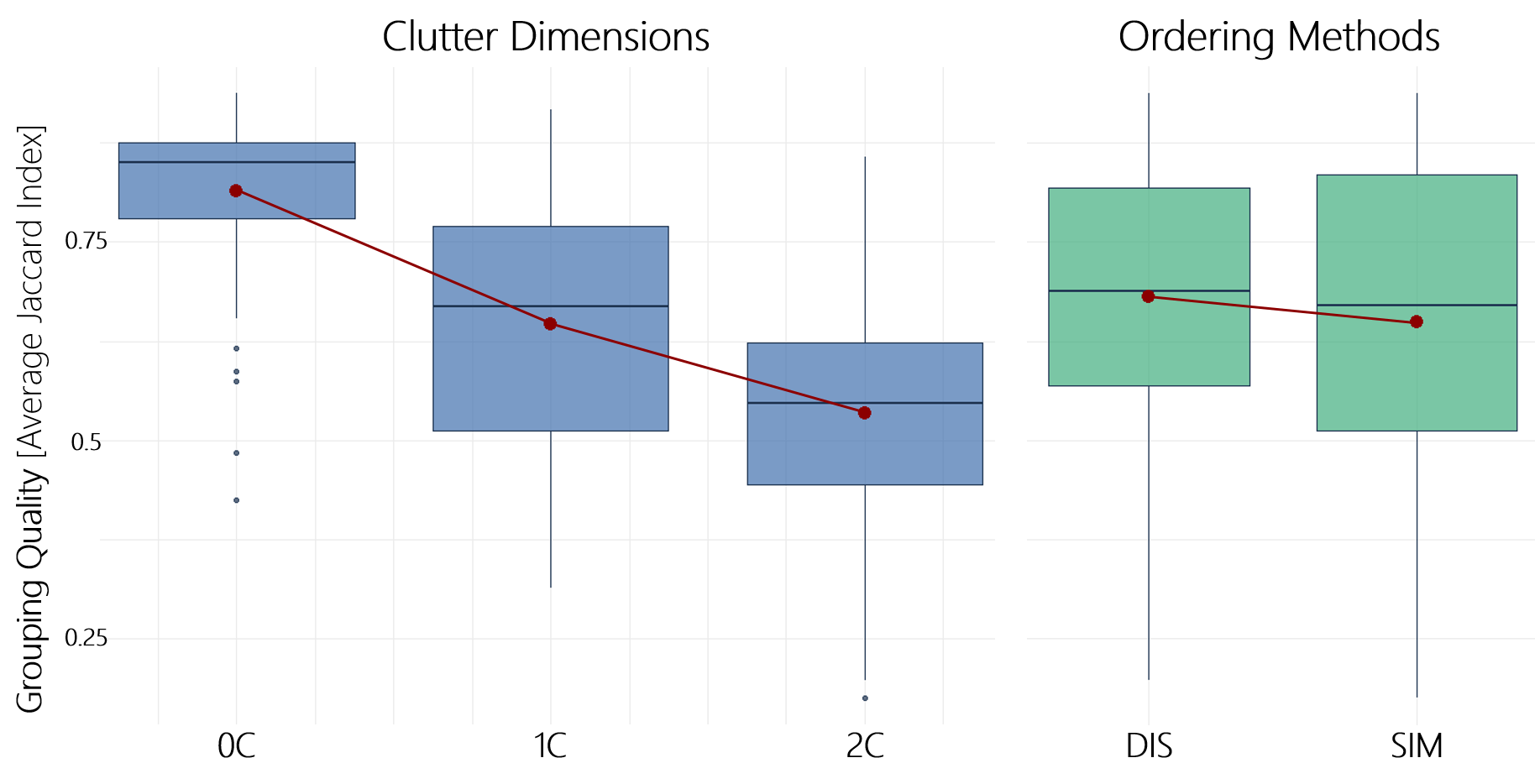}
    \vspace*{-0.25cm}
    \caption{\textbf{Cluster quality analysis.} Difference between clutter dimensions \noClutter, \oneClutter, and \twoClutter as well as \DIS and \SIM ordering.}
    \label{fig:quality-chart.png}
    \vspace*{-0.4cm}
\end{figure}

\subhead{Cluster Quality}.

\noindent An overview of the cluster quality is depicted in Fig.~\ref{fig:quality-chart.png}.

\noindent \textbf{H1b.} 
There were significant effects of \emph{clutter level} on \emph{cluster quality} ($\chi^2(2, N = 270) = 109.92, p < .001$).
Post-hoc tests showed a higher participants' accuracy with \noClutter ($\bar{x} = 0.85$) compared to \oneClutter ($.67$, $p < .001$) and \twoClutter ($.55$, $p < .001$), and between \oneClutter and \twoClutter ($p < .001$). 

\noindent \textbf{H2b.} 
When comparing \emph{ordering strategies}, participants were more accurate with \DIS ($\bar{x} = .69$) compared to \SIM ($\bar{x} = .67$, $p < .05$), which is also true within clutter levels \oneClutter and \twoClutter, but not \noClutter. 
\noClutter: \DIS ($\bar{x} = .85$, $\mu = .81$) vs. \SIM ($\bar{x} = .85$, $\mu = .82$), 
\oneClutter: \DIS ($\bar{x} = .68$, $\mu = .66$) vs. \SIM ($\bar{x} = .66$, $\mu = .63$), 
\twoClutter: \DIS ($\bar{x} = .58$, $\mu = .57$) vs. \SIM ($\bar{x} = .52$, $\mu = .50$, $p < .001$).

\subhead{Noise Identification Quality.}

\noindent \textbf{H1c.} 
There was a significant effect of \emph{clutter level} on \emph{noise identification} ($\chi^2(2, N = 270) = 80.02, p < .001$).
Post-hoc tests revealed that participants were more accurate with \noClutter ($\bar{x} = .8$) compared to \oneClutter ($\bar{x} = .5$, $p < .001$) and \twoClutter ($\bar{x} = .33$, $p < .001$).
In addition, there was a significant effect between clutter conditions \oneClutter and \twoClutter ($p < .001$).

\noindent \textbf{H2c.} 
In general, there is no difference between \SIM and \DIS w.r.t. noise identification quality (both $\bar{x} = .5$)
There are no differences for \noClutter (both $\mu = .77$, \DIS $\bar{x} = .88$, \SIM $\bar{x} = .8$) and \oneClutter (both $\bar{x} = .5$, $\mu = .52$). 
But for the \twoClutter clutter condition, there was also a significant effect of \emph{ordering strategy} on \emph{noise identification} ($t(44) = 2.18$, $p = .05$).
Participants working with \DIS were more accurate ($\bar{x} = .4$, $\mu = .39$) in comparison to \SIM ($\bar{x} = .33$, $\mu = .32$, $p < .05$).

\subhead{Confidence.} 

\noindent \textbf{H1d.} 
There was a significant effect of \emph{clutter level} on \emph{confidence} ($\chi^2(2, N = 270) = 28.816, p < .005$).
Post-hoc tests revealed a higher confidence with \noClutter ($\bar{x} = 2$) compared to \twoClutter ($1$, $p < .001$).

\noindent \textbf{H2d.} There is no significant effect between \SIM ($1$) and \DIS ($1$). While there is also no effect within the different clutter levels (\noClutter: $2 / 2$, \oneClutter: $1 / 1$, and \twoClutter: $1 / 1$), there seems to be a tendency that participants are more confident with \SIM without clutter dimensions and more confident with \DIS with increasing clutter. 

\subsection{Quantitative User Feedback}

\subhead{Ordering Preferences.}
11 out of 15 participants reported that they could see the clusters more clearly with dissimilarity reordering because they could use the orientation of the spikes as a determining factor. 
Some participants reported that they generally found the grouping tasks challenging, and they were not quite sure about the results. Interestingly, most of them said to have
personal preferences towards the patterns with more smooth and convex shapes, namely the patterns produced by similarity reordering.

\subhead{Similarity Estimation Strategies.}
The strategies reported by the participants can be grouped into three categories: 
(1)~the majority of participants focused primarily on the spikes' orientation; 
(2)~participants reported that they tried to find the center of a star glyph, and observe at which position of the glyph the center lies and how the gray area around the center is shaped; 
(3)~a few participants searched for unique shape-parts and matched it with others. 
\section{Discussion \& Future Work}

In summary, our study revealed two major findings.

\subhead{Clutter Analysis.}
Clutter negatively influences the visual comparison of star glyphs. 
There is a significant drop in cluster quality, noise identification quality, and confidence with an increasing number of clutter dimensions. 
Also, task completion time changed considerably, although not statistically significant. 
Therefore, we can partially confirm our hypotheses \textbf{H1a -- H1d}. 

We expected these results as more clutter hampers similarity estimation in clustering tasks.
As a result, cluster performance drops. 
While this is a general problem in information visualization~\cite{DBLP:series/isrl/2015-72}, it particularly affects star glyphs as clutter may change their shape considerably.  
Glyph designers should, therefore, think of using automatic algorithms to remove clutter dimensions, if possible. 

\subhead{Ordering Analysis.}
There are differences between the two evaluated ordering strategies. 
Generally, the quality of the clustering was significantly more accurate with \DIS, in particular for datasets containing clutter (\oneClutter, \twoClutter). 
Participants also performed the task slightly faster using \DIS. However, they were on average 10 seconds faster with \SIM in non-cluttered datasets. 
We can see that \DIS significantly supports noise identification for a cluttered dataset (\twoClutter), but we cannot see a difference for the other clutter conditions. 
While many participants reported that they prefer a dissimilarity-based layout, we cannot see a significant result from the study. 
However, analyzing the Likert-scale distributions reveal a tendency that participants are more confident with \SIM for clutter-free datasets (\noClutter) and with \DIS for cluttered datasets (\twoClutter). 
Across all trials, we can confirm the hypotheses \textbf{H2b} and \textbf{H2c}, but completion time (\textbf{H2a}) and confidence (\textbf{H2d}) depend on the properties of the dataset.

These results are in line with Klippel et al.~\cite{klippel_star_2009, DBLP:journals/cartographica/KlippelHLW09}. 
We found it interesting that the difference between \SIM and \DIS is even more striking in cluttered datasets. 
The spikes seem to help users in identifying clutter dimension and improving the overall clusters. 
However, we could also see that, without clutter, participants were faster and more confident using a similarity-based ordering. 
The remaining question is whether it would be possible to combine \SIM and \DIS into a combined ordering strategy. 
Our study did not reveal whether participants need as many spikes as possible or whether a few important spikes are enough to improve the cluster quality. Further research needs to be done in this area. 
Another relevant question is also how the rotation of entire glyphs influences grouping quality in clustering tasks and further investigation in this direction is advisable as, for example, already started by Fuchs et al.~\cite{DBLP:journals/tvcg/FuchsIBFB14}.


\subhead{Design Considerations.} 
With the results gained from our study, we derive the following design considerations: 

\noindent\textbf{(1)} As the performance of users drop considerably when clutter dimensions are present, glyph designers should try to \textit{avoid clutter by applying a feature selection} method first, if possible. \newline
\noindent\textbf{(2)} Since, for \textit{datasets with clutter}, salient shapes and spikes support grouping tasks, we recommend using \DIS \textit{strategies}. \newline
\noindent\textbf{(3)} For \textit{datasets without clutter}, we did not find a clear difference between \SIM and \DIS. As \SIM seem to be slightly faster and less error prone to rotation~\cite{klippel_star_2009, DBLP:journals/cartographica/KlippelHLW09}. We recommend to use this strategy.

\subhead{Limitations.} 
We identified two main threats to our results' validity. 

\noindent(1)~The number of trials (270) is rather small, in particular, for the effectiveness and efficiency analysis of a specific clutter level. 
This affects not only the statistical analysis, but outliers may also distort the results. 
The number of trials per participant cannot be increased with the current study design; otherwise, the study would take much longer than one hour. 
Therefore, we suggest repeating the study with more participants to increase the number of trials. 
(2)~While we designed our datasets with different cluster structures and distributions, we limited them by eight dimensions as Klippel et al.~\cite{klippel_star_2009}. 
There might be differences for datasets with less, more, or an odd number of dimensions. 

\section{Conclusion and Future Work}
We conducted an empirical user study to evaluate the impact of clutter and axes ordering to clustering performance with star glyphs. 
Our results show that users perform better when the glyphs represent salient shapes and spikes, which is achieved by a dissimilarity-based ordering of the dimensions. Furthermore, we elicited that there is a significant impact of clutter on the clustering performance in general.

As future work, we plan to extend and re-run the study based on our discussed limitations and include other reordering strategies, as well. 
Extending to that, we want to investigate whether there is an influence of the data characteristics and rotation (e.g., favor symmetrical glyph shapes) to the ordering strategy. 
If so, we are interested in developing techniques to select the most useful ordering strategy based on the given data and task. 
Finally, automatic ordering strategies should be compared to user-driven axes arrangements, which are determined by experts based on their domain knowledge.

\acknowledgments{
We thank David Pomerenke and the anonymous reviewers for their valuable feedback and support. 
Funded by the Deutsche Forschungsgemeinschaft (DFG, German Research Foundation) – Projektnummer 251654672 – TRR 161 (Project A03).
}

\bibliographystyle{abbrv-doi}

\bibliography{sg-study}
\end{document}